\documentclass[12pt]{article}
\usepackage[pdftex]{graphicx}
\begin{document}

\title{
{\Large \bf{Construction of coherent states for physical
algebraic systems}} }

\author{Y. Hassouni \\
Laboratoire de Physique Theorique,\\
Faculte des sciences-Universite Mohammed V-Agdal \\
Avenue Ibn Batouta B.P: 1014, \\
  Agdal Rabat Morocco. \\
e-mail: y-hassou@fsr.ac.ma \\
 e-mail: yassine@ictp.trieste.it\\
and\\
E. M. F. Curado and M. A. Rego-Monteiro \\
Centro Brasileiro de Pesquisas F\'\i sicas, \\
Rua Xavier Sigaud 150, 22290-180 - Rio de Janeiro, RJ, Brazil\\
e-mail: evaldo@cbpf.br and regomont@cbpf.br
}
\date{}
\maketitle
\begin{abstract}

\indent

    We construct a general state which is an eigenvector of the
annihilation operator of the Generalized Heisenberg Algebra. We
show for several systems, which are characterized by different
energy spectra, that this general state satisfies the minimal set of
conditions required to obtain Klauder's minimal coherent states.

\end{abstract}

\vspace{0.5cm}

\begin{tabbing}

\=xxxxxxxxxxxxxxxxxx\= \kill

\\


\end{tabbing}

\newpage

\section{Introduction}

Coherent states (CS) were introduced by Schroedinger in 1926
\cite{schrod} while he was studying the one-dimensional harmonic
oscillator system. The same mathematical objects, the coherent
states, were also studied by Glauber \cite{glauber} and by
Klauder \cite{ref1} four decades ago. Glauber found them
while he was studying the electromagnetic correlation function
\cite{glauber}. He also realized that these states
have the interesting property of minimizing the uncertainty
Heisenberg relation. Thus,
one could say that these states are the quantum states with the
closest behavior to a classical system.
CS have applications in many areas of physics \cite{review} and
since the birth of these states there has always been some interest
in investigating their algebraic properties \cite{review,novos}.

We point out that there is not a unique way to construct coherent
states. In fact, there are different approaches leading to them,
for instance, the well known Klauder \cite{kapproach} and
Perelomov-Gilmore approaches
\cite{perapproach}. In the first approach the coherent states are
constructed using the basis of the Fock representation of the
harmonic oscillator algebra, while in the second one this
construction is based on notions of group theory.

In our work, we deal with Klauder's approach which is
based on the construction of coherent states of the
Heisenberg algebra. This algebra appears in many areas of modern
theoretical physics and as an example we notice that the
one-dimensional quantum oscillator algebra is an
important tool in the second quantization approach.

Due to the relevance of Heisenberg algebra,
during the last two decades some effort has been devoted to
study possible deformations of the harmonic oscillator algebra
\cite{heisdeformado}. Along these years several groups have also
generalized the Heisenberg algebra (see for instance \cite{first}
- \cite{gha})\footnote{A GHA is not necessarily a deformed Heisenberg
algebra.}.
All these Generalized Heisenberg Algebras (GHA) are related among
them. In this paper we will use the GHA given in \cite{jpa}
just because in this version of the algebra, the Hamiltonian of
the physical
system under consideration belongs explicitly to the set of
generators of the
algebra, the other generators in this set being the step
operators of the system.

The version of the GHA given in \cite{jpa} is written
using a general function $f(x)$ called characteristic
function of the algebra which is connected with the energy spectrum
of the physical system under consideration.
It was shown in
\cite{comhugo} that there is a class of quantum systems
described by this GHA. This class is characterized by those quantum
systems having energy eigenvalues written as $\epsilon_{n+1} f(\epsilon_{n})$ where $\epsilon_{n}$ and $\epsilon_{n+1}$ are
successive energy levels and $f(x)$ is a different function for each
physical system.

In this paper, motivated by the procedure for constructing
the standard coherent states of the harmonic oscillator,
we build a state which is an eigenstate of the annihilation
operator of the GHA. In the main part of this paper we discuss
the circumstances this general vector state, eigenstate of
the annihilation operator of the GHA, satisfies the minimum set
of conditions required to construct Klauder's coherent states
for the following systems:
\begin{enumerate}
    \item harmonic oscillator,
    \item deformed harmonic oscillator,
    \item a general class of spectra\\
    and
    \item free particle in a square well potential.
\end{enumerate}

This paper is organized as follows. In Section II we summarize the GHA.
In section III we present a general expression for an eigenstate of the
annihilation operator of the GHA and show that this expression
satisfies the minimum set of conditions required to be Klauder's coherent
states for the cases enumerated above. In section IV we present
our conclusions.

\section{Generalized Heisenberg algebra}

Let us begin by reviewing the version of the GHA (\cite{first}
- \cite{gha}) given in \cite{jpa}. We stress once more that
all these Generalized Heisenberg Algebras (GHA) are related among
them. The version of the GHA we are going to review is described
by the generators $J_0, A, A^{\dagger}$ satisfying \cite{jpa}:

\begin{eqnarray}
J_0 A^{\dagger}  & = & A^{\dagger} f\left( J_0\right)
\label{adagger}\\
A J_0 & = &  f\left( J_0\right) A
\label{a} \\
\left[ A^{\dagger}, A\right] & = & J_0-f\left( J_0\right)
\label{aadagger}
\end{eqnarray}
where $A = (A^{\dagger})^{\dagger}, J_0=J_0^{\dagger }$ is the
Hamiltonian of the physical system under consideration and
$f\left( J_0\right) $
is an analytic function of $J_0$, called the characteristic function of the
algebra. A large class of type Heisenberg algebras \footnote{A type
Heisenberg algebra is an algebra having annihilation and creation operators
among its generators.}
can be obtained just by appropriately choosing the function $f(J_0)$.
The Casimir operator of this generalized algebra has the expression:

\begin{equation}
C=A^{\dagger} A - J_0 = A A^{\dagger} - f\left( J_0\right)\,
\label{casimir} .
\end{equation}
This algebra has a connection with the algebra independently proposed
in  \cite{gha},
where the authors introduced the Heisenberg algebra through the set of
elements $\left( a^{-},a^{+},I\right) $, satisfying
\begin{eqnarray}
\left[ a^{-},a^{+}\right]  & = & a^{-}a^{+}-a^{+}a^{-}\equiv \Delta ^{\prime } \, ,
\label{aamais}\\
\left[ a^{-},\Delta \right] & = & \Delta ^{\prime }a^{-} \, \, \,\,\,\,  \mbox{and}
\label{amenos}\\
\left[ \Delta ,a^{+}\right]  & = & a^{+}\Delta^{\prime} \label{amais} \, ,
\end{eqnarray}
with $\Delta =a^{+}a^{-}.$
The connection between (\ref{adagger}-\ref{aadagger}) and (\ref{aamais}-\ref{amais})
can be made by means of the simple identification:

\begin{eqnarray}
\Delta ^{\prime } & = & f\left( J_0\right) -J_0  \\
a^{+}  & = & A^{\dagger} \\
a^{-} & = & A \\
a^{+}a^{-} & = & J_0 \\
a^{-}a^{+} & = & f\left( J_0\right) \, .
\end{eqnarray}

Before starting the construction of the coherent states associated with
some physical systems by means of their related algebra, let us give a summary of its
representation theory: the $n$-dimensional irreducible representations of the algebras (1-3)
and (5-7) are given through the lowest eigenvalue of
$J_0$ with respect to the vacuum  state $\mid 0\rangle$:

\begin{equation}
J_0\mid 0\rangle =\alpha _0\mid 0\rangle \, .
\end{equation}
It is clear that for each value of $\alpha _0$ and for a set of
parameters of the
algebra (related to the function $f$), we have a different vacuum, all of them denoted here,
for simplicity, by $\mid 0\rangle $.
The solution of the representation theory problem is given in
\cite{jpa} for the linear and quadratic polynomials. The
$n$-dimensional representation theory is given through a general vector
$\mid m\rangle $ that is required to be an eigenvector of $J_0$,

\begin{equation}
J_0\mid m\rangle =\alpha _m\mid m\rangle \, ,
\end{equation}
where $\alpha_m = f^{(m)} (\alpha_0)$, the $m$-th iterate of $\alpha_0$ under
$f$, and under the action of $A$ and $A^{\dagger}$ we have:
\begin{eqnarray}
A^\dagger \mid m \rangle & = & N_m \mid m+1 \rangle
\label{adagon}\\
A  \mid m \rangle & = & N_{m-1} \mid m-1 \rangle
\label{aon}
\end{eqnarray}
where $N_m^2 = \alpha_{m+1} - \alpha_0$.

In \cite{jpa} it was showed that choosing for the characteristic
function of the GHA the linear function $f(x)=x+1$ the algebra in
eqs. (\ref{adagger}-\ref{aadagger}) becomes the harmonic
oscillator algebra and for $f(x)=q x+1$ we obtain in eqs.
(\ref{adagger}-\ref{aadagger}) the deformed Heisenberg algebra.
Moreover, it was showed in \cite{comhugo} that there is a class of
quantum systems described by these generalized Heisenberg
algebras. This class is characterized by those quantum systems
having energy eigenvalues written as
\begin{equation}
\epsilon_{n+1} = f(\epsilon_{n}) \, ,
\label{eq:class}
\end{equation}
where $\epsilon_{n+1}$ and $\epsilon_{n}$ are successive energy
levels and $f(x)$ is a different function for each physical
system. This function $f(x)$ is exactly the same function that
appears in the construction of the algebra in eqs.
(\ref{adagger}-\ref{aadagger}), which was called the characteristic
function of the algebra. In the algebraic description
of this class of quantum systems, $J_0$ is the Hamiltonian
operator of the system, $A^{\dagger}$ and $A$ are the creation
and annihilation operators. This Hamiltonian and the ladder
operators are related by eq.
(\ref{casimir}) where $C$ is the Casimir operator of the
representation associated to the quantum system under
consideration.

\section{Coherent states}

\indent Now, we are in position to build the coherent states
corresponding to some particular form of the characteristic
function. Let us construct a state $|z \rangle$ which is an
eigenstate of the annihilation operator of the GHA introduced
in the previous section, i. e.,
\begin{eqnarray}
    A |z \rangle = z |z \rangle,
\label{defcs}
\end{eqnarray}
where  $z$ is a complex number. We expand the state
$|z \rangle$ as $|z \rangle = \sum_{n=0}^{\infty}
c_n |n \rangle$. Acting the annihilation operator
of the GHA on $|z \rangle$ and using Eqs. (\ref{aon}) and
(\ref{defcs}) we have
\begin{eqnarray}
    A |z \rangle = \sum_{n=0}^{\infty} c_{n+1} N_n \,  |n \rangle
    = z \sum_{n=0}^{\infty} c_n |n \rangle .
    \label{constrcs}
\end{eqnarray}
Equating the coefficients of $|n \rangle$ gives $c_{n+1} N_nz c_n$. The solution of this equation for arbitrary $c_n$ is
\begin{eqnarray}
    c_n = c_0 \frac{z^n}{N_{n-1} !} ,
    \label{solcn}
\end{eqnarray}
where by definition $N_{n} ! \equiv N_0 N_1 \ldots N_{n}$
and by consistency $N_{-1} ! \equiv 1$. We will see in what
follows that this definition of $!$ reduces to the
standard definition of factorial for the harmonic oscillator
case. With the solution given in Eq. (\ref{solcn}) we obtain
for the state $|z \rangle$
\begin{eqnarray}
    |z \rangle = N(z) \sum_{n=0}^{\infty} \frac{z^n}{N_{n-1} !}
    \; |n \rangle,
    \label{propcs}
\end{eqnarray}
where we have used $N(z)$ instead of $c_0$.

Let us now recall what are the minimal set of conditions to
obtain Klauder's coherent states (KCS).

\noindent
A state $\mid z\rangle $ is called a KCS if it satisfies the following
conditions:

i/ Normalizability:

\begin{equation}\langle z\mid z\rangle =1
\end{equation}

ii/ Continuity in the label:

\begin{equation}
\mid z-z^{\prime }\mid \rightarrow 0;\,\, \,\,\, \parallel\, \mid z>-\mid z^{\prime
}>\parallel \rightarrow 0
\end{equation}

iii/ Completeness
\begin{equation}
\int d^2z \; w\left( z\right) \mid z\rangle \langle z\mid =1
\end{equation}

\noindent
We are going now to analyze the above minimal set of conditions
to obtain a KCS for the state given in Eq. (\ref{propcs})
in several examples.

\vspace{0.2cm}

\noindent
\underline{Harmonic oscillator}

\vspace{0.2cm}
As commented in the previous section, the GHA reduces to
the Heisenberg algebra by choosing the linear function
$f(x)=x+1$ for its characteristic function. In this case
we have $N_{n-1}^2 = n$ and Eq. (\ref{propcs}) becomes
the standard coherent state for the harmonic oscillator
with normalization coefficient given by
$N\equiv N\left(z\right) =exp\left( -\frac{\mid z\mid ^2}
2\right)$ and the weight function $w(r)$, $r=|z|$, required by
the third condition is $w(r) =\frac 1\pi $.

\vspace{0.2cm}

\noindent
\underline{Deformed Heisenberg algebra}

\vspace{0.2cm}

As discussed in \cite{jpa} by choosing the characteristic function
of the GHA as $f(x)=q x+1$ we obtain a deformed Heisenberg algebra.
In this case since $N^2_{n-1}=N_0^2 [n]_q$, where $N_0^2=\alpha_0
(q-1)+1$, the Gauss number being $[n]_q=(q^n-1)/(q-1)$ and $\alpha_0$
is the eigenvalue of the Casimir for the representation.

In the case we are analyzing Eq. (\ref{propcs}) becomes
\begin{eqnarray}
    |z \rangle = \frac{N(|z|)}{N_0} \sum_{n=0}^{\infty} \frac{z^n}{\sqrt{[n]_q !}}
    \; |n \rangle,
    \label{csqosc}
\end{eqnarray}
where $[n]_q !\equiv [1]_q\, [2]_q \, \ldots [n]_q$ and $[0]_q ! \equiv 1$.
Using the normalizability condition we have
\begin{eqnarray}
    |z \rangle = \left[ \sum_{n=0}^{\infty} \frac{|z|^{2n}}{[n]_q !}
    \right]^{-1/2} \sum_{n=0}^{\infty} \frac{z^n}{\sqrt{[n]_q !}}
    \; |n \rangle.
    \label{csqosc1}
\end{eqnarray}
As discussed in \cite{india} the function
$g(z)=\sum_{n=0}^{\infty} \frac{|z|^{2 n}}{[n]_q !}$
which appears in the above equation  is
convergent within a circle of radius $1/(1-q)$ for $0<q<1 $
and outside this
circle the function is defined by analytic continuation. For the
classical case (q=1) it was shown that the completeness condition
is achieved with $w(z)=1/\pi$ (see \cite{india} for details on the
construction of the weight function for this case).

\vspace{0.2cm}

\noindent
\underline{A class of spectra}

\vspace{0.2cm}

Let us now apply Eq. (\ref{propcs})  to a simple class of
spectra and then to the physically important case of the free particle in a square
well potential. The key point is to
know the analytical expression of the energy levels as shown below.

\vspace{0.2cm}
I/ spectrum type 1:

\vspace{0.2cm}
Let us consider a system whose spectrum is given by the
expression:
\begin{equation}
\varepsilon _n=1-\frac 1{n+1}=\frac n{n+1}, \, \,  \mbox{with}\, \,\,n\geq 0.
\label{eq:spectrum1}
\end{equation}
To obtain the characteristic function of the generalized algebra associated with this spectrum,
we remark that:
\begin{equation}
\varepsilon _{n+1}=\frac{n+1}{n+2}=\frac 1{\frac n{n+1}+\frac
2{n+1}} \,\, .
\label{eq:en1}
\end{equation}
As:
\begin{equation}
\varepsilon _n=\frac n{n+1} \,\,\,\,  \mbox{and} \,\,\,\,  1 - \varepsilon _n = \frac 1{n+1} \, ,
\label{eq:truque}
\end{equation}
the substitution of eqs. (\ref{eq:truque}) in eq. (\ref{eq:en1}) allows us to obtain the
recurrence equation:
\begin{equation}
\varepsilon _{n+1}=\frac 1{2-\varepsilon _n} \, \,\, .
\end{equation}
Thus,
\begin{equation}
\varepsilon _{n+1}=f\left( \varepsilon _n\right) =\frac 1{2-\varepsilon
_n} \,\, ,
\end{equation}
allowing us to identify the characteristic function $f$ to be used in the algebra
associated with this energy spectrum:
\begin{equation}
f\left( x\right) =\frac 1{2-x} \,\,\,.
\end{equation}
\noindent
As, for the present spectrum:

\begin{equation}
\alpha _n = \varepsilon_n  = \frac {n}{n+1} \, ,  \,\,\,\,\,\,\,\,\,          \alpha _0=0\,\,\, ,
\end{equation}
\noindent
it is thus easy to see that

\begin{equation}
N_{n-1}^2=\frac {n}{n+1}\,\, ,
\end{equation}
yielding

\begin{equation}
N_{n-1}!=\frac {1}{\left( n+1\right) ^{\frac{1}{2}}} \,\, .
\end{equation}
The vector $| z \rangle$ defined in Eq. (\ref{propcs}) can thus be written as:

\begin{equation}
\mid z\rangle =N\left( \mid z\mid ^2 \right) \sum_{n \ge 0} \left( n+1\right) ^{\frac
{1}{2}}z^n\mid n\rangle \,\, .
\label{eq:kcs}
\end{equation}

Now, following our proposal concerning the definition of KCS, we have to
verify the three conditions mentioned above.
Requiring that $ \langle z\mid z\rangle = 1$
(normalizability condition), and remembering that $\langle m | n \rangle = \delta_{m,n}$,
one obtains,
\begin{equation}
\label{ }
 \langle z\mid z\rangle = N^2 (|z| ^2) \sum_{m \geq 0} (m+1) |z|^{2 m} \,\, .
\end{equation}
As:
$$
\sum_{m \geq 0}  (m+1) |z|^{2 m} = \frac{1}{(1-|z|^2)^2} \,\, ,
$$
we have for the normalization factor:
\begin{equation}
N^2 \left( \mid z\mid ^2 \right) =(1-|z|^2)^2 \,\,,
\label{eq:norma}
\end{equation}
where $0 \leq |z| < 1$.
Let us remark that with this result, obtained from a
particular spectrum, the KCS can be constructed with a normalization function
that is different from the exponential function which is the standard case.
\noindent
The second condition (continuity condition) is automatically verified. But
to satisfy the third one which is, in general, the most important, we have to
to find the weight function allowing the equality:

\begin{equation}
\int d^2z \, w\left( \mid z\mid ^2\right) \mid z\rangle  \langle z\mid =1 \,\, .
\label{eq:third}
\end{equation}
This expression means
the over completeness condition in the KCS
domain for the particular case of the spectrum of type 1.
Substituting eqs. (\ref{eq:kcs}) and (\ref{eq:norma}) in eq. (\ref{eq:third}) and
integrating on the angle $\theta$ ($z = r \exp(i \theta)$), we
obtain the expression:
\begin{equation}
\label{ }
2\pi \sum_{m\ge 0}\mid m\rangle \langle m\mid \int_0^1 dr \, w(r^2) N^2 (r^2) \left( m+1\right) r^{2m+1} \, .
\end{equation}
Changing $x=r^2$, this expression can be written as:
\begin{equation}
\label{ }
\pi \sum_{m\ge 0}\mid m\rangle \langle m\mid \int_0^1 dx \, w(x) N^2 (x) \left( m+1\right) x^m \, ,
\end{equation}
and remarking that:

\begin{equation}
\int_0^1dx\left( n+1\right) x^n=1\,\, ,
\end{equation}
it s obvious that we can solve eq. (\ref{eq:third}) if we
choose the weight function $w$ satisfying the
condition:

\begin{equation}
\pi w\left( x\right) N^2 \left( x\right) =1 \,\, .
\end{equation}
The explicit form of $w$, allowing the resolution of the
completeness equation can, finally, be written as:

\begin{equation}
w(x)=\frac {1}{\pi (1-x)^2 } \,\,\, .
\end{equation}

II/ spectrum type 2:

\vspace{0.2cm}
Now, we are going to treat the case of the quadratic spectrum in this class
of spectra. Let us call quadratic spectrum the spectrum $\varepsilon _m$
having the following
expression:

\begin{equation}
\varepsilon _m=\left( 1-\frac 1{m+1}\right) ^2=\frac{m^2}{\left( m+1\right)
^2} \,\, ,
\end{equation}
where $m = 0, 1, 2, 3, \dots $.
As in the previous case, we are interested to compute the
characteristic function of the GHA for this particular spectrum.
>From the above expression, we have:

\begin{equation}
\sqrt{\varepsilon _m}-1=\frac{-1}{m+1} \,\,\, ,
\end{equation}
leading us to the expression:

\begin{equation}
\varepsilon _{m+1}=\frac{\left( m+1\right) ^2}{\left( m+2\right) ^2}\left( \frac 1{2-
\sqrt{\epsilon _m}}\right) ^2 \,\,\, .
\end{equation}
Consequently, the characteristic equation is given by:

\begin{equation}
\varepsilon _{m+1}=f\left( \varepsilon _m\right)\,\, ,
\end{equation}
with

\begin{equation}
f\left( x\right) =\left( \frac 1{2-\sqrt{x}}\right) ^2 \,\,\, .
\end{equation}
As before, let us consider Eq. (\ref{propcs}).
As $\alpha_0 =  \varepsilon_0 =  0$ and  $N_{m-1}^2 = \alpha_{m} - \alpha_0 = \alpha_{m}
= \varepsilon_m  = (m/(m+1))^2$,
after some calculation, the scalar $\langle z\mid z\rangle $ can be written as:

\begin{equation}
\label{ }
\langle z\mid z\rangle =N^2 \left( \mid z\mid ^2 \right) \sum_{m \ge 0}\left(
m+1\right) ^2\mid z\mid ^{2m} \,\, .
\end{equation}
The sum can be easily performed
and we obtain:

\begin{equation}
N^2 \left( \mid z\mid ^2 \right) \frac{(1-|z|^2)^3}{1+|z|^2} \,\, ,
\label{eq:n2}
\end{equation}
where $0 \leq |z| <1$.
We note that once more the normalized function is not an exponential one.

As mentioned before the most important equation is the resolution of the
completeness equation. To get this, we must find an adequate weight function $w$.
Performing a computation similar as in previous cases,
the weight function must obey ($x = | z | ^2$):

\begin{equation}
\pi \sum_{m\ge 0}{\mid m\rangle \langle m\mid }
\int_0^1dx \, w\left( x\right) N^2(x)\left( m+1\right) ^2x^m=1 \, .
\end{equation}
One can verify that a solution of this equation is given by:

\begin{equation}
w\left( x\right) =-\frac{\ln x}{\pi N^2\left( x\right) } \,\,\, .
\end{equation}
Using eq. (\ref{eq:n2}), we can write the weight function as:

\begin{equation}
w\left( x\right) =-\frac{\ln x}\pi \frac{1+x}{(1-x)^3} \,\, .
\end{equation}
This function ensures the resolution of the completeness equation, corresponding
to the case of the spectrum of type 2, allowing the construction of coherent
states for this kind of spectrum.

\vspace{0.2cm}

III/  general case
\vspace{0.2cm}

The spectra of types 1 and 2 can be
generalized to an arbitrary order. Let us now consider the general spectrum:

\begin{equation}
\varepsilon _n=\left( 1-\frac 1{n+1}\right) ^\alpha ,
\end{equation}
with $\alpha \geq 2$. the question now is to find the corresponding
GHA. After that, we have to find the characteristic function.
Starting from the fact that:

\begin{equation}
\varepsilon _n^{1/\alpha }-1=-\frac 1{n+1} \,\, ,
\end{equation}
one can check that:

\begin{equation}
\varepsilon _{n+1}=\left( \frac{n+1}{n+2}\right) ^\alpha =\left( \frac
1{1-\varepsilon _n^{1/\alpha }}\right) ^\alpha \,\, .
\end{equation}
Then the characteristic function is:

\begin{equation}
f(x)=\left( \frac 1{2-x^{1/\alpha }}\right) ^\alpha \,\, .
\end{equation}
Let us now verify the minimal set of conditions for the state in
Eq. (\ref{propcs}) in the case of
the general spectrum under consideration.
For this general case we have:

\begin{equation}
N_{n-1}!=\frac 1{\left( n+1\right) ^{\frac \alpha 2}} \,\, ,
\end{equation}
allowing us to write Eq. (\ref{propcs}) in this case as:

\begin{equation}
\mid z\rangle =N\left( \mid z\mid ^2 \right) \sum_{n \ge 0} \left( n+1\right) ^{\frac
\alpha 2}z^n\mid n\rangle \,\, .
\end{equation}
As
\begin{equation}
\label{ }
\sum_{n \ge 0} \left( n+1\right) ^{\alpha} |z|^{2n} = \frac{Li_{-\alpha} (|z|^2)}{|z|^2} \,\, ,
\end{equation}
where $Li_{-\alpha} (|z|^2)$ is the polylogarithm function,
the coefficient $N^2 (\mid z\mid ^2 )$ can be written as:

\begin{equation}
N^2 (\mid z\mid ^2 )=\frac{|z|^2}{Li_{-\alpha} (|z|^2)} \,\, ,
\end{equation}
and $0 \leq |z| < 1$.
The expression of the weight function that is behind the resolution
of the unity equation is, nevertheless, harder to be obtained.
Following the method used before, we find that,
for the general case,  the weight function can be written as
($x = |z|^2$):

\begin{equation}
w\left( x\right) =\left( -\right) ^{\alpha +1}\frac{\left( \ln x\right)
^{\alpha -1}}{\pi \Gamma(\alpha) N^2 \left( x\right) } \,\,\, ,
\end{equation}
where $\Gamma(\alpha)$ is the gamma function.
\noindent
As an example, we consider the behavior of $w(x)$ for $\alpha =3$:

\begin{equation}
w\left( x\right) = \frac{\left( \ln x\right)
^{2}}{ 2 \pi} \frac{1+ 4 x + x^2 }{(1-x)^4} \,\, ,
\end{equation}
which is shown in Figure 1.

\vspace{0.2cm}
\noindent
\underline{Free particle in a square-well potential: }
\vspace{0.2cm}

We are going now to compute the coherent states of a physical
system using the formalism described before.
The results in this case are more complicated because
they involve a spectrum needing a weight function which is
a special function.
In fact, the latter is relatively not obvious in comparison with
the ones introduced in the previous sections.
Let us begin with the well-known spectrum of a free-particle in a square-well potential:

\begin{equation}
\varepsilon _n=(n+1)^2 \,\, , \,\,\, \,\,\,\,\,\,\,\,\,\,\,\,  n = 0,1, 2, 3, \dots
\end{equation}
Then,

\begin{equation}
\varepsilon _{n+1}=\left( n+2\right) ^2 = \varepsilon_n + 2 \sqrt{\varepsilon_n}  + 1 \,\, .
\end{equation}
Using the algebraic formalism shown before and observing that ($\alpha_n = \varepsilon_n$),

\begin{equation}
\label{eq:Nfree}
N_{n-1}^2=\alpha_n - \alpha_0 = \alpha_n - 1= n^2+2n \,\, ,
\end{equation}
we obtain,

\begin{equation}
N_{n-1}! = \frac 1{\sqrt{2}}\sqrt{n!}\sqrt{\left( n+2\right) !} \,\, .
\end{equation}
We thus obtain for our proposal of coherent states given in Eq. (\ref{propcs})
the expression:

\begin{equation}
\mid z\rangle =\sqrt{2}\, N(\mid z\mid )\sum_{n \ge 0} \frac{z^n}{\sqrt{n!}\sqrt{\left(
n+2\right) !}}\mid n\rangle .
\end{equation}
The normalizability condition can be fulfilled if we satisfy the expression:

\begin{equation}
2N^2 \left( \mid z\mid \right) \sum_{n \ge 0} \frac{\mid z\mid ^{2n}}{n! \left(
n+2\right) !}=1 \,\, .
\end{equation}
Noting that:
\begin{equation}
\sum_{n \ge 0} \frac{|z|^{2n}}{ n! \, ( n+2)!} = \frac{ I_{2}(
2\mid z\mid ) }{ \mid z\mid ^2} \,\, ,
\end{equation}
for $0 \leq |z| <1$,
where $I_n (z)$ is the modified Bessel function of the first kind of order $n$,
the expression for the normalizability coefficient can be written as:
\begin{equation}
N^2 \left( \mid z\mid \right) =\frac{\mid z\mid ^2}{2 I_{2}(
2\mid z\mid )  } \,\, ,
\end{equation}
where $0 \leq |z| < \infty$.
The behavior of this function can be seen in Figure 2.
The resolution of the completeness problem is given by finding the adequate weight
function $w(x)$, $x=|z|^2$, satisfying the equality:

\begin{equation}
\pi \sum_{n\ge 0}{\mid n\rangle  \langle n\mid \frac
2{n!(n+2)!}}\int_0^{\infty }dx\, \frac{ w\left( \sqrt{x} \, \right)
 N^2(\sqrt{x}\, ) x^{n+1}}{2x}=1 \, .
\end{equation}
If we take:

\begin{equation}
\frac{\pi w\left( \sqrt{x} \, \right) N^2 (\sqrt{x}\, ) }{2x}
=K_2 \left(2\sqrt{x}\right) \,\, ,
\end{equation}
where $K_n (x)$ is the modified Bessel function of the second kind
of order $n$, the weight function takes the form:

\begin{equation}
w\left( \sqrt{x}\, \right) =\frac 2\pi \frac{xK_2 \left( 2\sqrt{x}
\right) }{N^2 \left( \sqrt{x} \right) } \,\, ,
\end{equation}
and can, finally, be written as:

\begin{equation}
w\left( x\right) =\frac 4 \pi K_2 \left( 2 \sqrt{x}\right)  \,  I_{2}(
2 \sqrt{x} )  ,
\end{equation}
which is shown in Figure 3.
With this expression, one can verify that the important condition,
the completeness equation, is satisfied by considering that:

\begin{equation}
\int_0^{\infty }dx \, K_2 \left( 2\sqrt{x} \, \right) x^{n+1}=\frac 12 n! \left(
n+1\right) ! \,\,\,.
\end{equation}

\vspace{0.2cm}

\section{Conclusion}

We have investigated in this work a state constructed as an
eigenstate of the annihilation operator of the Generalized
Heisenberg Algebra (GHA). We have shown for several systems
(harmonic oscillator, deformed harmonic oscillator, a class of
spectra and the square well potential) that this state satisfies the
minimum set of conditions required to obtain Klauder's coherent
states.

The GHA we considered, is an algebra having as generators the
Hamiltonian of the physical system under consideration and the
annihilation and creation operators of the system. The state we
have investigated is an eigenstate of the annihilation operator
of the GHA for a general system described by this GHA. Thus, this
state is a natural generalization for a general system described
by the GHA of the coherent states of the standard harmonic
oscillator system.

It is interesting to note that
in the proof (Eqs. (\ref{defcs}-\ref{propcs}))
of our expression  for coherent states given in Eq. (\ref{propcs})
it was only necessary to admit i) an infinite sum, ii) $A |0\rangle=0$ and
iii) $A |n\rangle = N_{n-1}|n-1\rangle$. Thus, the explicit expression
of $N_{n}$ was not necessary in order to get Eq. (\ref{propcs}). The
explicit expression of $N_{n}$ was necessary only when we showed, for
specific spectra, that the state satisfied the minimal set of conditions
to obtain Klauder's coherent states. Thus, we think that the expression
in Eq. (\ref{propcs}) could be a consistent definition of coherent
states even for systems which are not described by the GHA but
satisfying the conditions (i-iii) mentioned above.

\vspace{0.4 cm}

\noindent
{\bf Acknowledgments:}
E. M. F. Curado and M. A. Rego-Monteiro thank CNPq/Pronex for
partial support. Y. Hassouni thanks M. El Baz for discussions
and TWAS/CNPq for partial support.

\vspace{0.4 cm}
\noindent
{\bf Figure Captions:}
\vspace{0.2 cm}
\newline
Figure 1: Weight function for the system having the
characteristic function given as $f(x)=(1/(2-x^{1/3}))^{1/3}$.
\vspace{0.2 cm}
\newline
Figure 2: Normalization function for the free particle in a
square well potential.
\vspace{0.2 cm}
\newline
Figure 3: Weight function for the free particle in a square well
potential.


\begin{thebibliography}{99}

\bibitem{schrod} E. Schroedinger, Naturwissenscahften
{\bf 14} (1926) 664.

\bibitem{glauber} R. J. Glauber, Phys. Rev. {\bf 131} (1963)
2766.

\bibitem{ref1} J. R. Klauder, J. Math. Phys. {\bf 4} (1963)
1058.

\bibitem{review} See for instance: J. R. Klauder and
B. Skagerstam, Coherent States: Applications in physics and
mathematical physics, World Scientific, 1985, Singapore;
W-M Zhang, D. H. Feng and R. Gilmore, Rev. Mod. Phys. {\bf 62}
(1990) 868.

\bibitem{novos} See for instance: T. Shreecharan, P. K. Panigrahi
and J. Banerji, Phys. Rev. {\bf A 69} (2004) 012102; C. Quesne,
J. Phys. { \bf A 35} (2002) 9213; B. I. Lev, A. A. Semenov, C. V.
Usenko and J. R. Klauder, Phys. Rev. {\bf A 66} (2002) 022115;
S. Nouri, Phys. Rev. {\bf A 65} (2002) 062108;
G. S. Agarval and J. Banerji, Phys. Rev. {\bf A 64}
(2001) 023815.


\bibitem{kapproach} J. R. Klauder and B. S. Skagertan, Coherent
states, Singapore, World Scientific (1985).

\bibitem{perapproach} A. M. Perelomov, Commun. Math. Phys.
{\bf 26} (1972) 222.

\bibitem{heisdeformado} A. J. Macfarlane, J. Phys. { \bf A 22}
(1989) 4581; L. C. Biedenharn, J. Phys. { \bf A 22}
(1989) L415.

\bibitem{first} C. Quesne and N. Vansteenkiste, J. Phys.
{\bf A 28} (1995) 7019.

\bibitem{caos} E. M. F. Curado and M. A. Rego-Monteiro,
Phys. Rev. {\bf E 61} (2000) 6255.

\bibitem{jpa} E. M. F. Curado and M. A. Rego-Monteiro,
J. Phys. {\bf A 34} (2001) 3253.

\bibitem{gha} M. El Baz, Y. Hassouni and F. Madouri, Rep.
Math. Phys. {\bf 50} (2002) 263.

\bibitem{comhugo} E. M. F. Curado, M. A. Rego-Monteiro and
H. N. Nazareno, Phys. Rev. {\bf A 64} (2001) 12105; hep-th/0012244.

\bibitem{india} T. K. Kar and G. Ghosh, J. Phys. {\bf A 29}
(1996) 125.

\end{thebibliography}
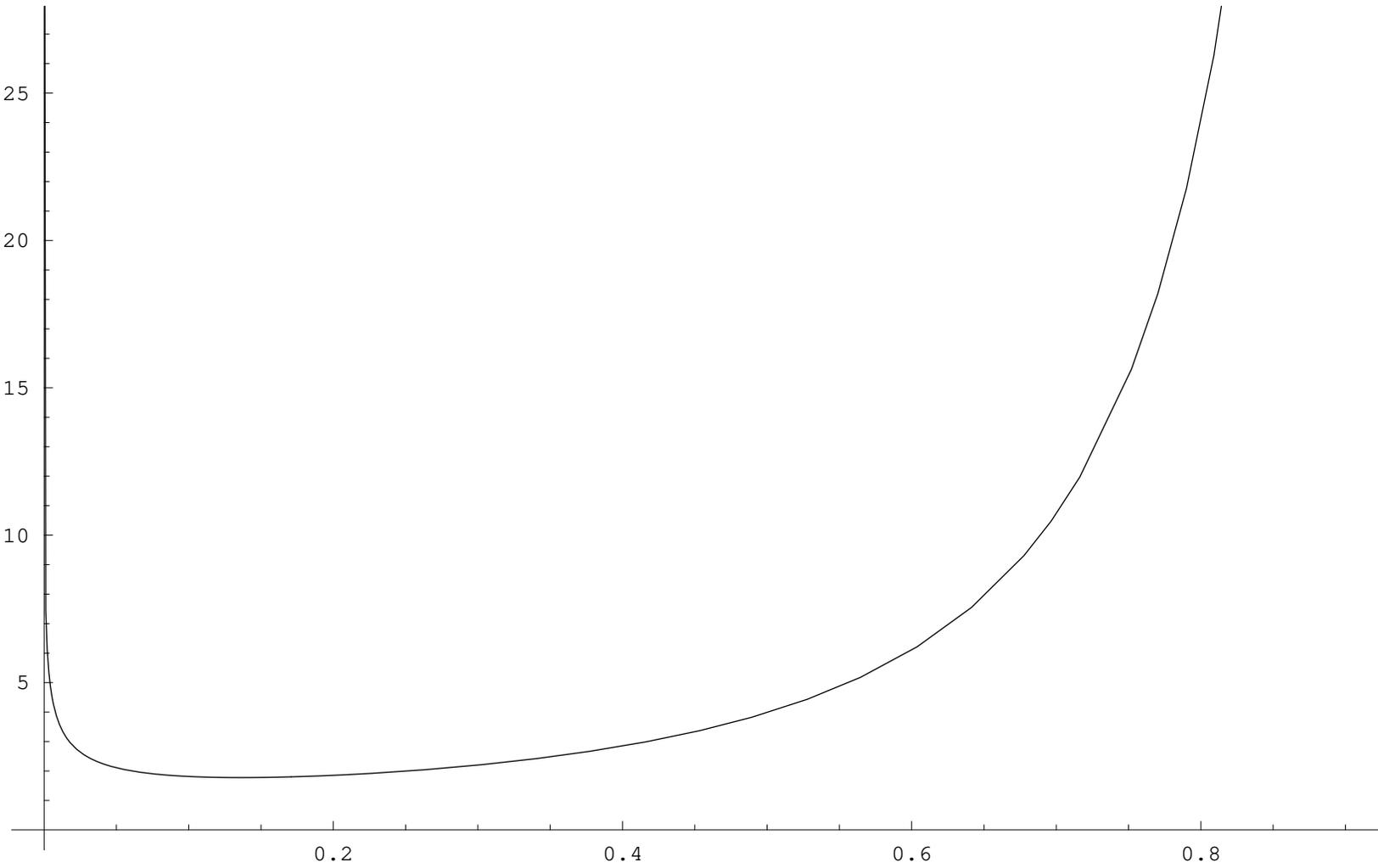
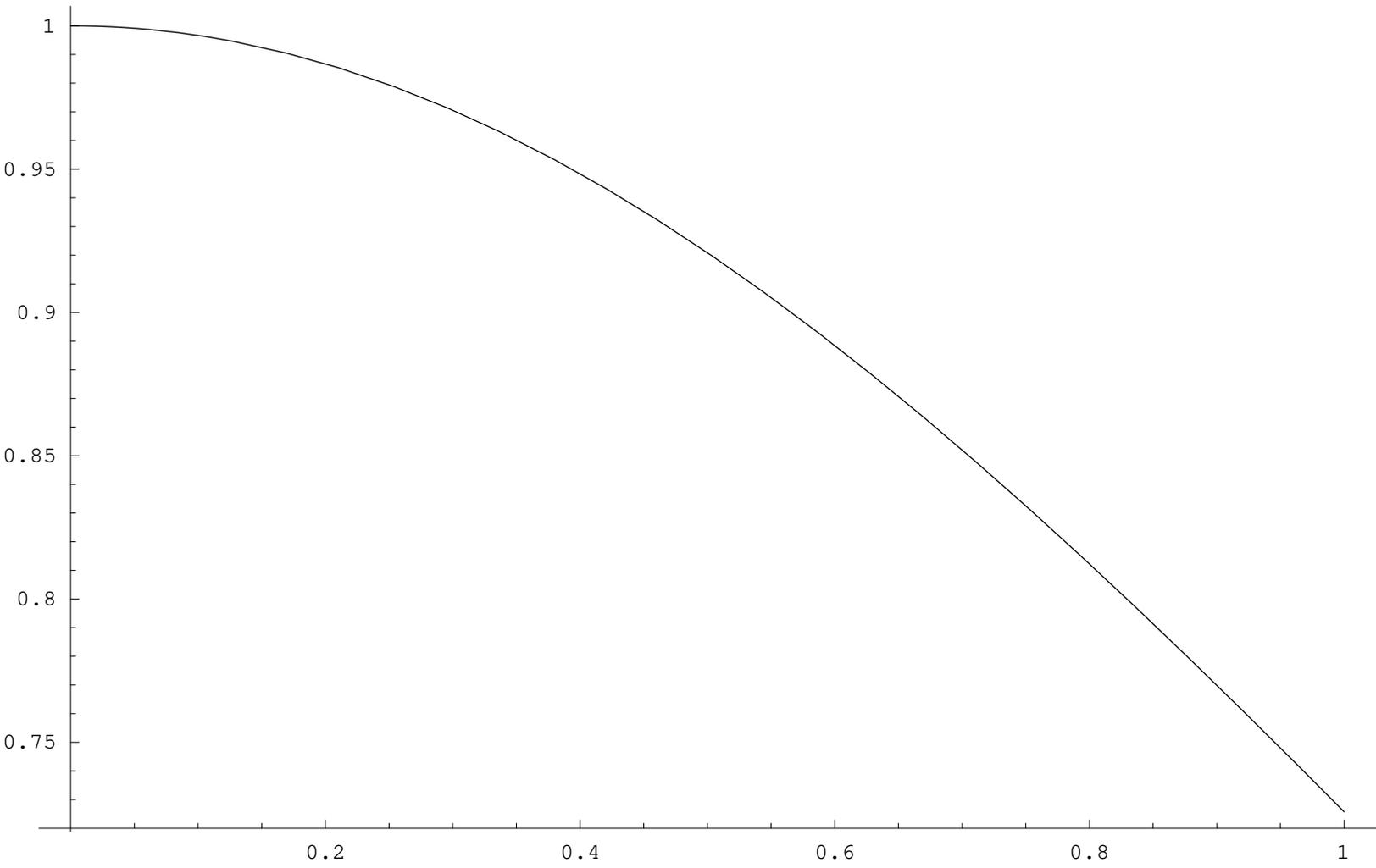
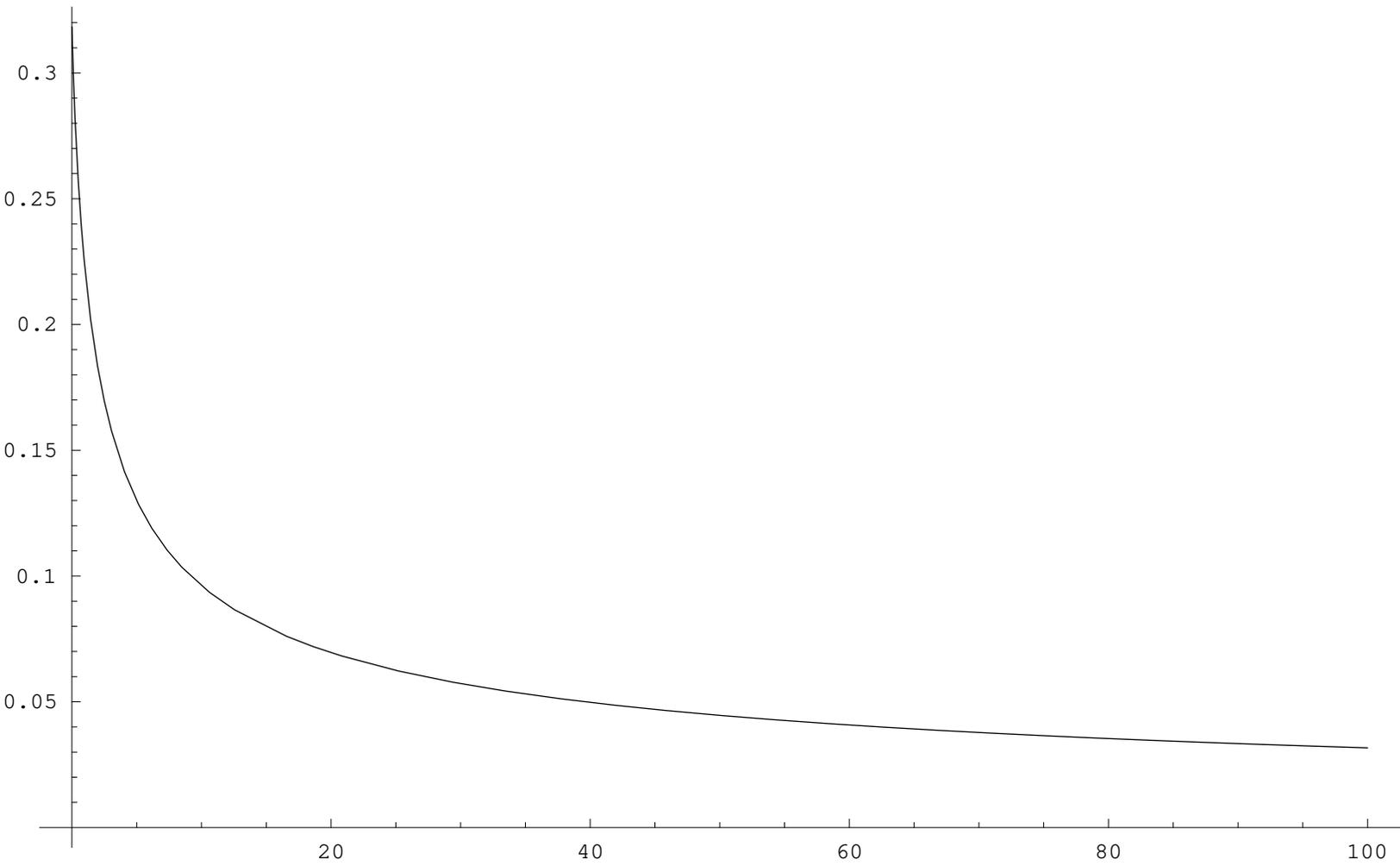
\end{document}